# Diffusion model and analysis of diffusion process at lagrangian method.

W2005430531366897104

ZHOU Ziqi[1] SUN Zhongguo[2] ZHANG Kai[3]

**ABSTRACT** Based on Fick's 2nd law the development of moving particle semi-implicit method for predicting diffusion process is proposed in this study. The diffusion model described how mass get transported between particles and the density of particles get fitted in real time. By giving the optimal radius for Laplacian operator, the accuracy of MPS diffusion solver can be optimistically estimated. A more accurate characterization method is provided by the diffusion model. Droplets collision and moving boundary were simulated and two internal flow cases were performed with concerns of the distribution of concentration in the fluid domain and the outlet sections. Factors affecting diffusion process were obtained. Analysis in unsteady flow at Lagrangian view was taken by extracting the pathlines and varying curves along the pathlines.

# Introduction

The Latin word 'diffundere' is the origin of diffusion which means 'to spread out'. Diffusion is caused by the Brownian motion of atoms or molecules that leads to complete mixing. Even water molecules in pure water are still in continuous random motion. The migration through the liquid itself is called 'self-diffusion'. The Scottish chemist Thomas Graham (1805-1869) [1]performed the first systematic studies of diffusion, giving a now call Graham's law: 'The diffusion or spontaneous intermixture of two gasses is effected by an interchange in position of indefinitely minute volumes of the gases, which volumes are not of equal magnitude, being, in the case of each gas, inversely proportional to the square root of the density of that gas.' In the field of studying diffusion, an important advance came from German physiologist Adolf Eugen Fick (1829-1901). Fick postulated that the flux of matter j in x direction is proportional to the pertaining gradient of concentration C [2]: $j = -D\frac{\partial C}{\partial x}$, which we nowadays call Fick's first law. By using the conservation of matter in analogy to Fourier's treatment, the second fundamental law of diffusion was given: $\frac{\partial C}{\partial t} = D\frac{\partial^2 C}{\partial x^2}$, which we nowadays call Fick's second law or diffusion equation. The diffusion equation is a linear partial differential equation of second order mathematically. Best known as physicist, Albert Einstein derived a relation [3] between the diffusivity of particles suspended in a liquid and the solvent viscosity η. With the Stokes friction force, $6\pi\eta r$, a relation for solute molecules of radius r was given: $D = \frac{R_g T}{N_A}\frac{1}{6\pi\eta r}$, the Stokes-Einstein relation so called nowadays.

Marvin Appel [4] at 1968 give solutions generalized to apply to an n-phase system, and emphasized that these solutions are only valid if the principal physical process is diffusion. C.R.Houska and J. Unnam[5] once gives the concentration distribution for any continuous variation of the diffusion coefficient, which required the diffusion geometry to be planar and the system to be doubly infinite in size, then use a finite difference approach to determine the variation of composition with time and distance. C.Jacoboni [6] studied electron diffusion in semiconductors at short distance and time .From both analytical and numerical approach to study the diffusion phenomenon at high fields and small distances and short time. E.J.Mittemeijer and H.C.F. Rozendall [7] proposed a rapid method for numerical solution of Fick's second law where the diffusion coefficient is concentration dependent. This rapid method is two orders of magnitude faster than before and should therefore be preferred in practice. W.E. Alley and B.J. Alder [8] suggested that Fick's law must be modified by replacing the diffusion coefficient by a convolution over time or the velocity autocorrelation function. Jeremiah R.LOWNEY [9] concluded that Fick's law is the more fundamental and straightforward way to model diffusion processes. A.J. Slavin and P.R. Underhill indicated that Fick's first law is also unchanged by explicitly including the occupancy of potential diffusion sites in the derivation. R.N. Hills and P.H. Roberts [10] drew attention to a disturbing inconsistency in some theoretical arguments that have been used to determine the evolution of regions of mixed solid and liquid phase. [11] L. Botar studied the migration and reaction-driven diffusion of two component and treated as a random walk on a simple cubic lattice. It is shown that in this model the cross-term in Fick's second law vanish even if the gradients of the concentration are not zero. J.M. Rubi et[12], discussed inertial effects in systems outside equilibrium within the framework of non-equilibrium thermodynamics. The elements of the theory of internal degrees of freedom were used to constitute the mesoscopic version of a previous analysis which considers the kinetic energy of diffusion. M. Howard Lee [13] succeed to establish the validity of Fick's law precisely from a microscopic theory. David Jou and Jose Casas-Vazquez [14] constructed a higher-order hydrodynamics for material motion in fluids, under arbitrary nonequilibrium condition. Stephen W. Webb and KARSTEN Pruess [15] conbined gas-phase diffusion and advection in porous media, the advective-diffusive model (ADM) and the dusty-gas model (DGM). A modified Klinkenberg factor is suggested to account for difference in the models. I.V. BELOVA et [16] simulated a direct steady state using computer simulation method to calculate Onsager phenomenological transport coefficient from the gradient of the chemical potential in the one component lattice gas .B Ph van Milligen [17]focus on Fick's law and Fokker-Planck law and re-examine the origin of these expressions and perform numerical and physical experiments to shed light on this duality. Tomozo SASAKI [18] doubt that there is no guarantee that Fick's law holds true in soils. The authors think it is imperative to confirm the validity of Fick's law before starting measurements of radon diffusion coefficients in soils to obtain the confidence in the results. ZHANG Junzhi [19] used Fick's second law and water-cement ratios or porosity of concrete and chloride concentration to predict the varying law of chloride diffusion coefficient with exposure time of existing concrete. Chiara Visentin [20]and Nicola Prodi investigated the validity of the Fick's law of diffusion in room acoustics. Siepmann et[21] studied modeling of diffusion controlled drug delivery. S. Guenneau and T. M. Puvirajesinhe [22] introduced a coordinate transformation approach to control diffusion processes via anisotropy with an emphasis on concentration of chemical species for potential application in biophysics or bioengineering.

Gilberto M. Kremer[23] proposed that the coefficients of diffusion depend on the gravitational potential and become smaller than those in its absence. S. Chen et [24] proposed a new nonlinear two-side space-fractional diffusion equation with variable coefficients from the fractional Fick's law and a semi-implicit difference method for this equation.

Fick's law gives path to describe the mass transfer phenomenon between materials. CFD techniques were developed to solve much more complicated cases with mathematical tools, and numerical analysis was taken when experiment is difficult to sample and monitor especially when the simulating domain varies with respect of time. Considering the time-varying domain an important factor for CFD calculation, mesh methods (FVM, FEM and FDM et) had to paid more effort (dynamic mesh) to equip that. Species Transport Model is developed for mesh method to calculate the diffusion problem. If the domain is time-varying or free surface or moving boundaries were involved, Species Transport Model needs the support of dynamic mesh to provide a new analytical domain and updates boundary condition at every time step. Moreover, Eularian description includes a convection term causing difficulties of solving and numerical dissipation.[25] According to the difficulties and inconvenience, solving Fick's second law the linear partial differential equation in particle method or called as meshless method could be more appropriate. In this paper, mass diffusion process is directly incorporated in the MPS method[26]. By redefining the particles as fluid physical variables' storage domain instead of considering particles as simply single fluid particle, the particles now denote sub-domains of fluid. Extending the attributes of particles with the concentration, one particle can describe an assemble of several component matters sharing the information of coordinate, velocity, pressure and other physical variables. This study includes building the solving method of Fick's 2$^{nd}$ law and validating the accuracy of the Laplacian operator. The calculated cases were to introducing the potential of diffusion model and application of boundary conditions. The analysis gives a clear understanding about how unsteady flow affects the diffusion process.

The core techniques of MPS method are given below and the newly proposed diffusion model is also given as follow.

## Moving particle semi-implicit (MPS) method

### Governing equations

Continuity equation:

$$\frac{D\rho}{Dt}+\rho\nabla\cdot\boldsymbol{u}=0 \qquad (1)$$

The incompressible Navier-Stokes equation:

$$\rho \frac{D\boldsymbol{u}}{Dt} = -\nabla p + \mu \nabla^2 \boldsymbol{u} + \rho \boldsymbol{g} + \boldsymbol{f}_s \tag{2}$$

Where $\boldsymbol{u}$ denotes the velocity, t denotes the time, $\rho$ denotes the density, $p$ denotes the pressure, $\mu$ denotes the kinetic viscosity coefficient, $\boldsymbol{g}$ denotes the acceleration of gravity, $\boldsymbol{f}_s$ denotes surface tension translated into a force per unit fluid volume.

## Discretization

Kernel function calculates the interaction between two particles:

$$w(r) = \begin{cases} \dfrac{r_e}{r} - 1 & (r \le r_e) \\ 0 & (r > r_e) \end{cases} \tag{3}$$

$r_e$ denotes the effective radius of interaction

The particle number density $n_i$ of a fluid particle $i$ is defined as follows:

$$n_i = \sum_{j \ne i} w(|r_i - r_j|) \tag{4}$$

where $r = |r_i - r_j|$ is the distance between particle $i$ and particle $j$

Ensuring $n_0$ constant during solving process achieve the incompressibility of fluid.

## Gradient and Laplacian operators

Gradient operator and Laplacian operator in the MPS method:

$$\langle \nabla \varphi \rangle_i = \frac{d}{n^0} \sum_{j \ne i} \left[ \frac{\varphi_j - \varphi_i}{r_{ij}^2} (\boldsymbol{r}_i - \boldsymbol{r}_j) w(|\boldsymbol{r}_i - \boldsymbol{r}_j|) \right] \tag{5}$$

$$\langle \nabla^2 \varphi \rangle_i = \frac{2d}{n^0 \lambda} \sum_{j \ne i} [(\varphi_j - \varphi_i) w(|\boldsymbol{r}_i - \boldsymbol{r}_j|)] \tag{6}$$

where $d$ is the number of space dimensions

Parameter $\lambda$ is introduced to make the increment of variation equal to the analytical solution:

$$\lambda = \frac{\sum_{j \neq i} w(|\mathbf{r}_i - \mathbf{r}_j|) |\mathbf{r}_i - \mathbf{r}_j|^2}{\sum_{j \neq i} w(|\mathbf{r}_i - \mathbf{r}_j|)} \cong \frac{\int_v w(r) r^2 dv}{\int_v w(r) dv} \tag{7}$$

Use these models, the gradient and Laplacian of physical quantity $\varphi$ on particle $i$ can be obtained as the weighted summation of quantity $\varphi$ on particles in its effecti radius with a weight function $w$.

## Boundary condition

### Surface particle detection

In the MPS method, a free-surface particle of a fluid field satisfies:

$$n_i < \beta n_0 \tag{8}$$

Where $n_i$ denotes particle number density, $\beta$ is set as 0.97 to ensure the stability of calculation.

The pressure of the free-surface particle is set to zero Pa as a boundary condition in pressure calculating.

### Wall boundary condition

Wall boundary conditions can be described as: (1) no-slip and (2) slip. In this paper, the velocity of the wall particles is set to 0 m/s and the viscosity of the wall particles is set to the viscosity of the liquid to reproduced the no-slip wall boundary condition.

### Velocity Inlet and Pressure outlet boundary conditions

The inlet and outlet boundary conditions were formed before explicit and implicit computations. Therefore, when calculating the domain, it ensures that boundary conditions work at the time step. Reclaim the boundary conditions at every time step, so the continuity can be guaranteed. Before explicit and implicit computations, particles are to be added near the inlet boundaries, called the candidate particles. The coordinates of candidate particles are calculated as $r_{cand} = r_{bc}\boldsymbol{n}_{bc}d_{add}$, where $r_{cand}$ denotes the coordinate of candidate particle, $r_{bc}$ is the coordinate of boundary, $\boldsymbol{n}_{bc}$ is the normal vector, $d_{add}$ is the selected distance. Do collision detection near $r_{cand}$ at a certain distance (mostly $1.01l_0$). If pass, make up particles near the boundaries and assign the physical value to them. For outlet boundaries, the particles get removed from the computation domain. The removing domain is given by stretch the boundary lines (surfaces in 3D) at the direction of normal vector with distance $d_{outlet}$. When particles move into the removing domain, the angle of the $v_i$ and $n_{bc}$ is calculated by $v_i \times n_{bc}$. If it shows particle $i$ is approaching to the outlet boundaries, then the particle $i$ get removed from the computation domain, otherwise the particle stays. For inlet and outlet boundary conditions, the key factor to simulate them is to hold given condition at every time step. The velocity inlet condition is given by reclaiming the velocities of particles near the inlet boundary. The same the pressure outlet is also given by reclaiming the pressure of particles near the outlet boundary at every time step. It can be extended to any physical variable. The momentum input of the system is well-controlled in this method and the stability will not decrease.

## Diffusion model

Fick's second law predicts how diffusion causes the concentration $C$ to change with respect to time. The partial differential equation reads:

$$\frac{\partial C}{\partial t} = D\nabla^2 C \tag{9}$$

In the limit of low Reynolds number, diffusion coefficient $D$ can be described by Stokes-Einstein relation:

$$D = k_B T / 6\pi\eta r \tag{10}$$

where $k_B$ denotes the Boltzmann constant, $T$ is the absolute Temperature, $\eta$ is the viscosity,

$r$ is the spherical particles' radius

Concentration of components could be described as $C = f(x, y, z, t, T)$, in which $x, y$ and $z$ are the Eulerian location, $t$ is time and $T$ is the local temperature. In Lagrangian, description was given as $C = f(i, t, T)$, in which $i$ is marked number of particles. In that case, particle $i$ could be considered as a mesoscopic fluid group made up of components. The information of possibility of any component was carried by particle $i$ at microscopic view, while at macroscopic the information can be altered with $C$ the concentration. In this condition, a new diffusion method was proposed to predict how diffusion causes the concentration to change with respect to time applied to particle method.

Approximately Solving the Laplacian $\nabla^2 C$ in MPS reads:

$$\nabla^2 C = \frac{2d}{\lambda n^0} \sum_{j \neq i} (C_j - C_i) w(|r_j - r_i|) \qquad |r_j - r_i| \leq R_{local} \qquad (11)$$

$$\nabla^2 C = 0 \qquad |r_j - r_i| > R_{local} \qquad (12)$$

$\frac{\partial C}{\partial t}$ discrete as $\frac{C^* - C}{dt}$

With Stokes-Einstein relation, we obtain

$$C^* = C + \frac{\nabla^2 C \times k_B T}{6\pi\eta r} \times dt \qquad (13)$$

## Physical property fitting

With the diffusion advancing, particle density varies with respect to concentration. Consequently, those physical property terms were modified according to the variation of certain substance's concentration. Practical solution to density was given as follows

$$\rho_f = \frac{c_1\rho_1 + c_2\rho_2}{c_1 + c_2} \tag{14}$$

# Validation

2-D diffusion is a classic problem in computational solution of Fick's 2$^{nd}$ Law. The validation case was set as two component diffusion in a 2-D square space. Concentration was represented with normalized c, where $c = 1$ indicated the particle is fully occupied by component a and $c = 0$ means the particle is fully occupied by the component b in two component system. In this case, the distribution of concentration can be obtained by analytical solution of unsteady mass diffusion:

$$c_x = c_1 + (c_{0.5} - c_1)\left(1 - \text{erf}\left(\frac{-x}{2\sqrt{Dt}}\right)\right) \tag{15}$$

Giving $c_1$=1, $c_{0.5}$=0.5 and D=0.00001. The results were calculated at t=0.01s.

The accuracy of Laplacian operator varies depending on giving different value of $R_{local}$, the radius of neighborhood particle searching. Giving too high, the local detail will be lost resulting process accelerated. Giving too small, the stochasticity takes advantage, resulting calculation unreliable. Based on that, the optimal $R_{local}$ is tested in the case of two component diffusion to select reasonable value. $R_{local}$ to be tested includes 2.1l0(most used), 2.5l0, 2.9l0, 3.1l0, 4.0l0. The concentration distribution is given as fig. 1 at t=0.01s with $R_{local} = 2.5l0$.

By comparing the distribution in fig.1, result of 2.5l0 shows most agreement to the analytical solution. The basic tendency is also obvious to obtained: The larger the $R_{local}$ is, the faster (refer to the analytical solution) the process gets. Learn from fig.1, $R_{local}$ is set as 2.5l0 in next few cases.

The validation of Laplacian operator with different sampling time at $R_{local} = 2.5L0$ was obtained by fig.2.

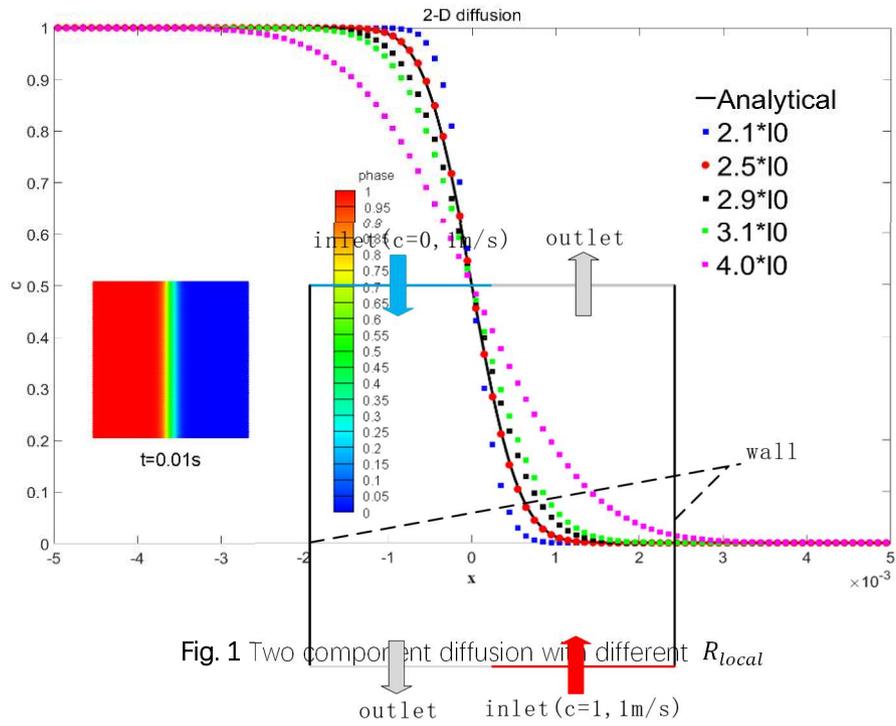

Fig. 1 Two component diffusion with different $R_{local}$

Fig.3 Boundary conditions arrangement

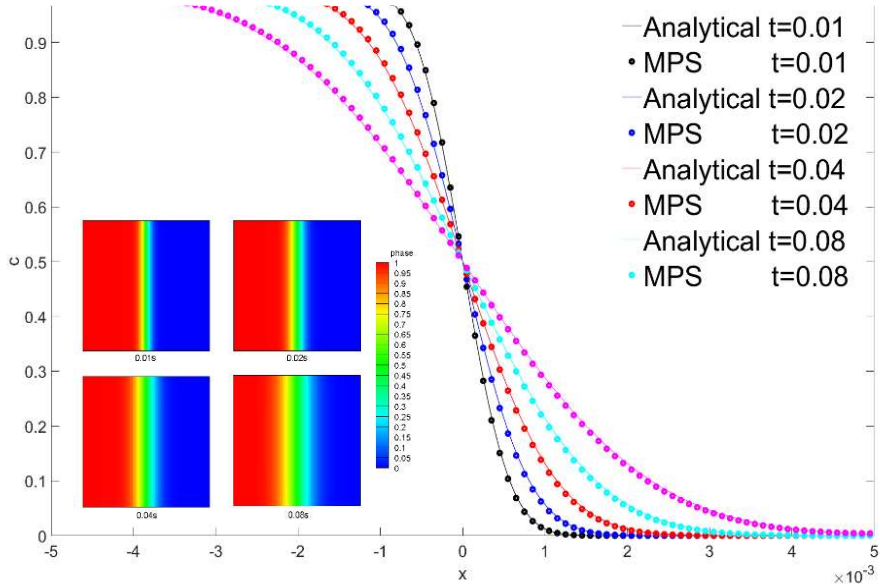

Fig.2 Concentration distributions comparison at different time

It indicates that the Laplacian solver has almost the same accuracy regardless the selected sampling time. Further on the case above, the center symmetrical boundary conditions were applied to calculate how diffusion advances when flow is included. The boundary conditions were as follow:

The case is also calculated in Ansys Fluent with Species Transport Model. The totally the same boundary conditions were set as table 1. The viscous model was chosen as Laminar and properties of fluid were defined identically. The case was performed as D=0.00001 with transient solver at a 100×100 construct mesh.

The MPS results were post-processed by triangulation. As Fig.4(a) shows, the central part of the concentration distribution was not vertical as fig.1 shows but a slope due to the flow working. The results of MPS solver share the same characteristics with the results of Fluent (the mesh-based solver). Furthermore, the concentration of distribution, velocity distribution and streamline all compare well with the results of Fluent.

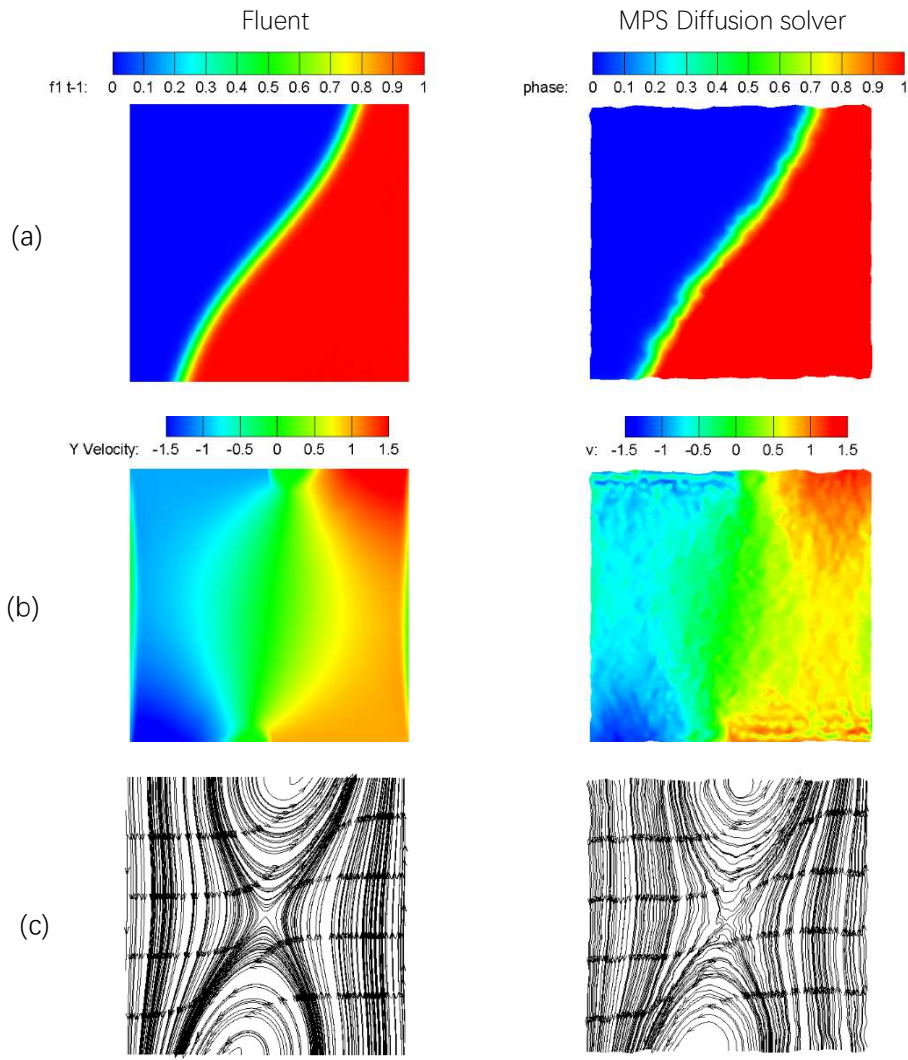

Fig.4 Comparison of 2-D diffusion with flow. (a): the concentration distribution; (b): the velocity in Y direction; (c: the streamlines)

# Results and discussion

In our study, we will discuss free surfaces, moving boundaries, steady-state internal flow, and unsteady internal flow.

## Head-on droplet collision

One of advantages of MPS method is solving free surface problem[27]. Fig 5 was performed as the collision of two different component droplet. In this case, the Diffusion solver is coupled and particle density was auto fitted at every time step. By giving a diffusion coefficient D=0.00001, density of upper droplet: 1000kg/m³ and density of droplet below: 2000 kg/m³. The droplets were given velocities of opposite 10m/s, gravity was not involved and surface tension was working in the Fig.6 gives how concentration and centroids varies with respect of time. The droplets merged and mixed after collision. At first, the interface got enlarged, then due to surface tension, the merged droplet retracted to the spindle shape under the effect of surface tension, forcing flow and giving help to accelerate diffusion process. The density of lower droplet is twice as the upper one, not only causing the merged droplet moving upward after full momentum exchange, but also leading to the concentration center part (around c=0.5) an upward offset.

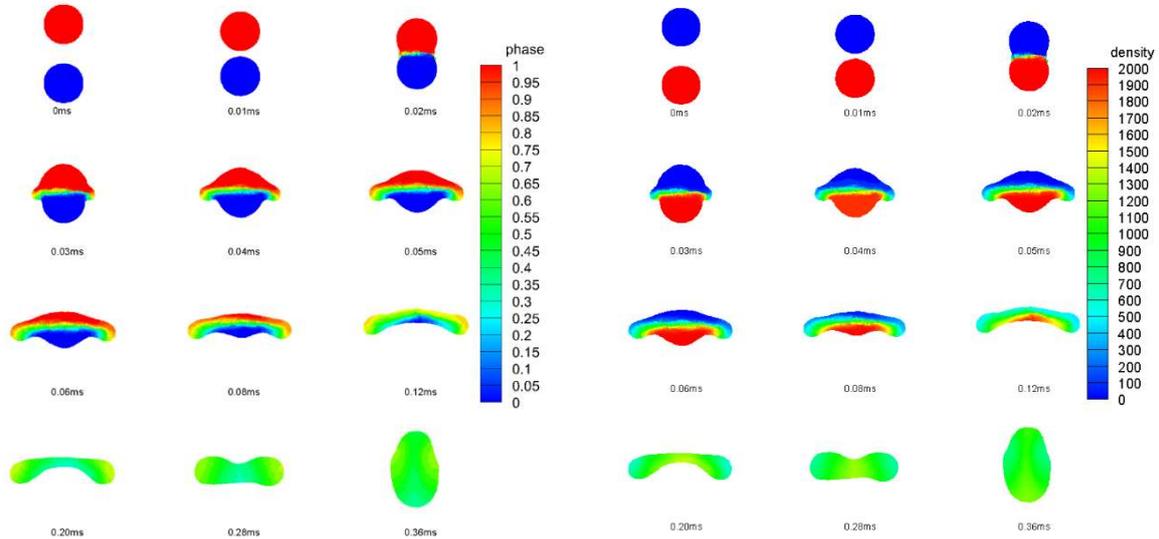

Fig.5 Collision of different droplet (the views of sequent illustration were manually centered for better presentation, in fact the center of gravity of droplets is moving upward due to the momentum work; left: concentration, right: density)

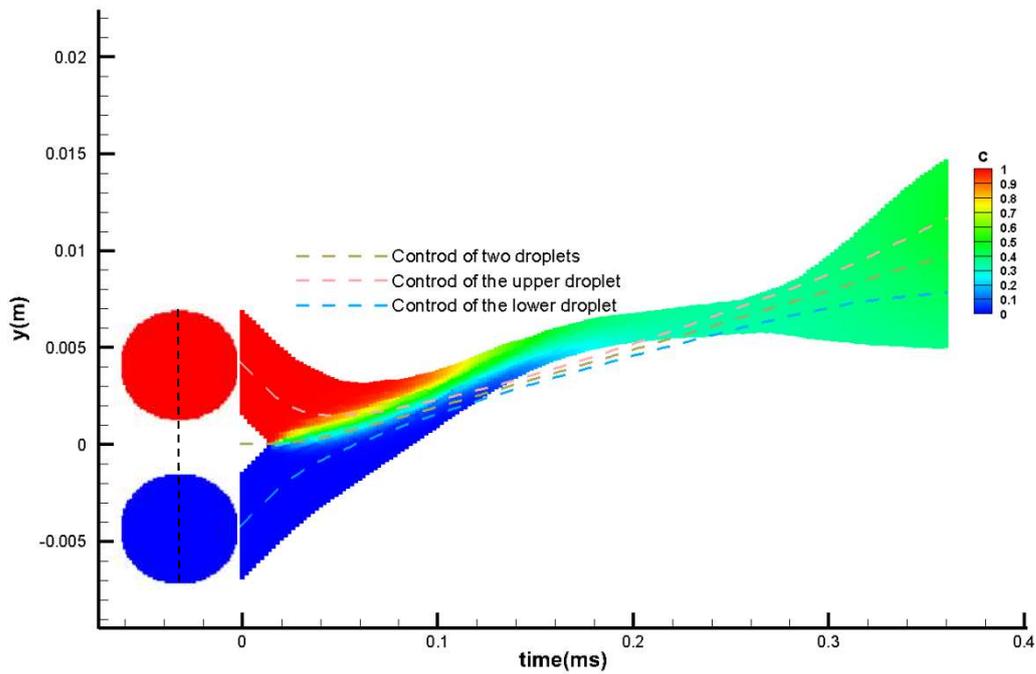

Fig.6 Concentration slice map (x=0: black dashed line) and centroid position

## Moving boundary

Moving boundary is used to simulate problems with fluid-solid coupling[28]. A Circular active mixer(radius:R=0.1m) was set as fig.x shows, the cylinder at diameter:R/5 rotates on radius:R/2 with line speed: $V_l$ =1m/s. The diffusion coefficient is set to 0.00001.

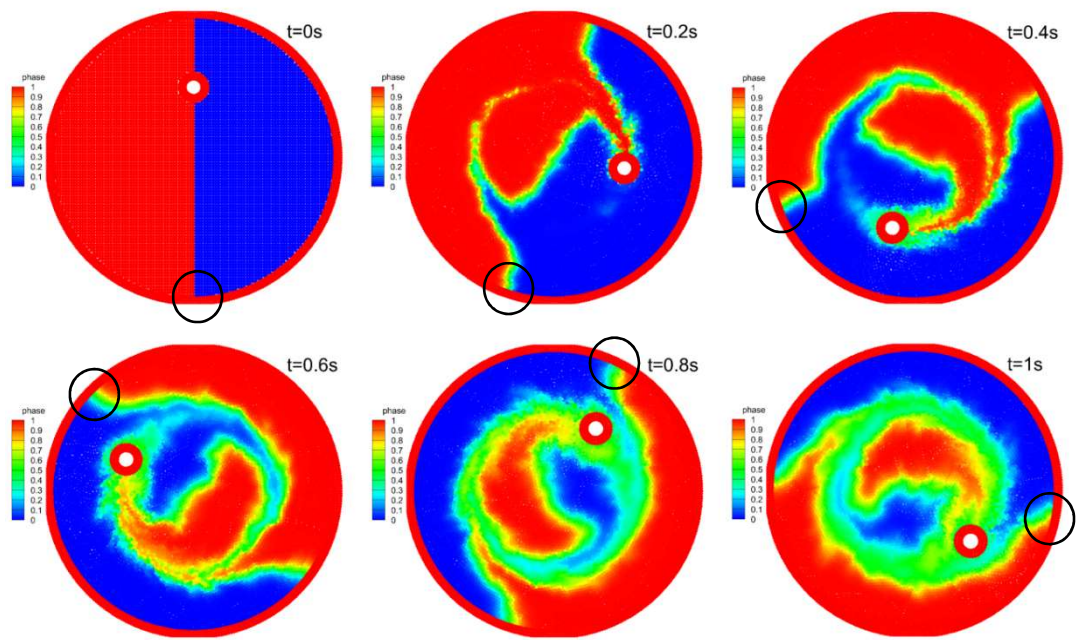

**Fig.7** Sequence illustration of concentration distribution

Fig 7 show the sequence illustration in 1 second. Due to the angular momentum input, the fluid so possesses a rotation movement as circled in fig.7. With the mixing advancing, the trailing area of cylinder was well diffused. Because the continues rotation movement of cylinder left the trailing area filled with high concentration gradient fluid as fig.7 t=0.2, continually enlarging the interface area. According to fig.7, the interface finally divided the whole domain into five parts: A1, A2(the low concentration); B1, B2(the high concentration) and C (transition region) as illustratde in fig.8.

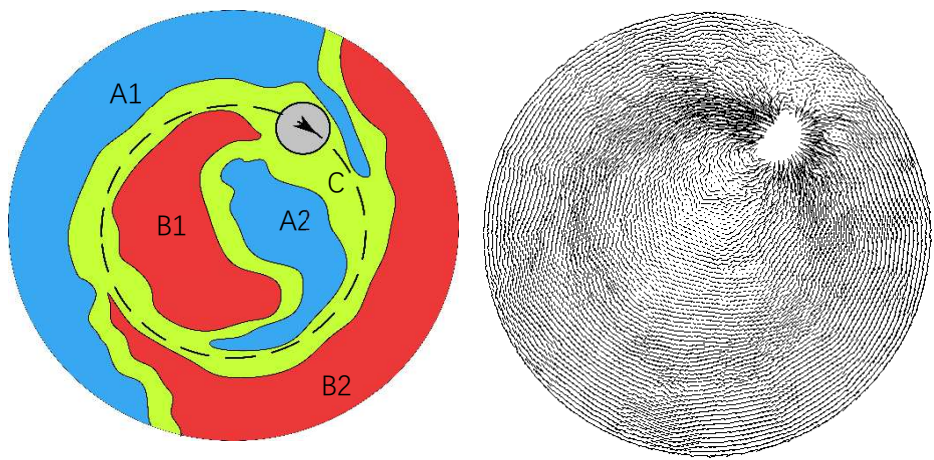

Fig.8 Illustration of fluid parts (left) and vector map (right)

The vector map reveals how the fluid flow during rotational mixing. Bernoulli formula tells that the trailing area will result in a negtive pressure area and from the vector map, it is clearly that the fluid from both side: A1 and B1 or A2 and B2 would strongly mix at the trailing area. Base on that, the help handed by mixing could be considered as two: enlarging and bringing effect. The enlarging effect happens from the initial moment untill the fluid parts were separated as fig.9 show; while the bringing effect works as always, however it became the main help only when the fluid parts were formed.

# Internal flow

Internal flow is the typical problem where boundary conditions work as the key factors. simulations of separated inlet tunnel and two-component combined in one inlet flow were taken to analysis the process.

## Extension of 2-D square diffusion

When the four sides were applied with similar boundary conditions, the central part of the concentration distribution was not a slope as fig.10 shows. It was extended to an area as indicates:

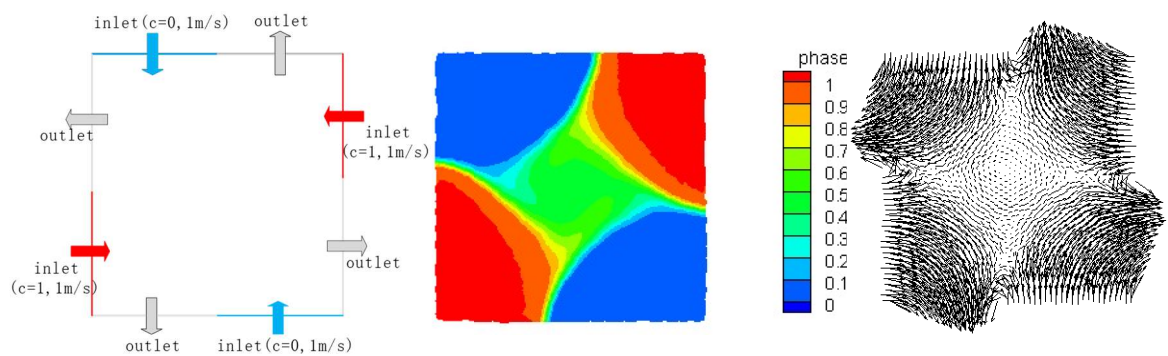

Fig. 10 Contour and vector of four pair boundary conditions

The contour of concentration has the property of central symmetry, result from the central symmetrical application of boundary condition. The vector map indicates the quasi-stagnation status of flow in center part.

## Separated inlet tunnel (FIG12 未提)

Fig.11 gives two typical geometries of symmetrical inlet tunnel or the 'T tunnel' and the side inlet tunnel. The velocity boundary conditions were applied to the virtual inlet line (the red and the blue) and the pressure outlet boundary condition was also applied to the outlet line (the yellow). The so-called side inlet tunnel exchanged the places of one inlet and the outlet. The flow rate ratio of two inlet was set as 1:1 in symmetrical inlet tunnel and one inlet size was get halved to change the ratio to 1:2 in the side inlet tunnel. The Reynold number was 50 where the laminar flow dominates. The T tunnel was empty as fig.11 shows at the initial moment. With calculation advancing the computation domain will auto fill with particles. Fig.13 gives the distribution of outlet line in a duration of 0.005s by scattering the particles' concentration to the coordinate plane. The center-symmetrical curve results from symmetrical inlet boundary condition. The concentration varies with along the horizontal direction. Obviously, the distribution gets more even and the fluctuating band shortens when increasing the diffusion coefficients. The symmetrical center locates the middle of outlet line resulting in the concentration of $c = 1/2$, while the center locates at the $c = 1/3$ in the side inlet tunnel. The basic regularities are the same with results of symmetrical inlet, while the property of symmetry get deprived due to the vertical inlet. The distribution center locates about c=1/3 and the fluctuations greatly increase and gradually decrease from the lower wall to the upper wall most obviously at the diffusion coefficient D=0.0005 as fig 13 shows.

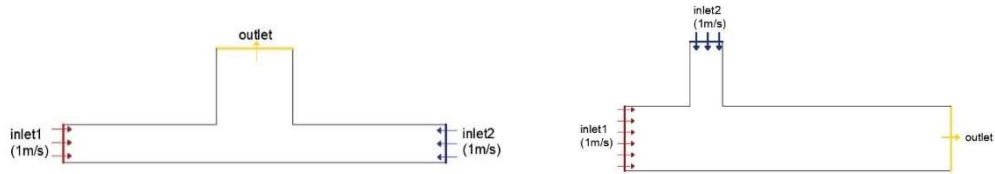

Fig.11 Geometries and boundary conditions

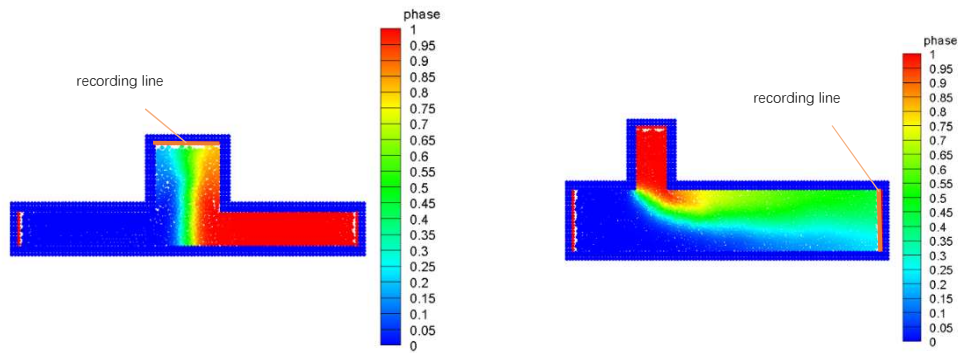

Fig.12 Contour of concentration distribution when flow is steady

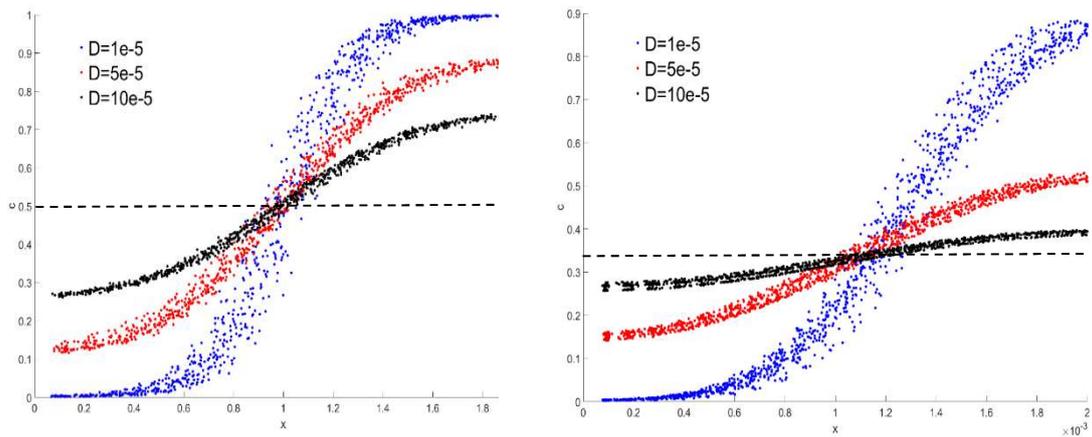

Fig.13 Scattering distribution of concentration

## Two-component flow around a cylinder

When the diffusion progress needs to meet the demand of resulting in more even distribution, it is necessary to make shear flow and try lowing the flow stability. In that case, a cylinder at radius=D was placed in a 2D tunnel at length=L and width=H. The D/H is called the blockage ratio, the case below was set as D/H=0.5 and the center of cylinder was place at 1/3 of the length of tunnel. The inlet was divided into two halves, the upper half produce the first component and lower half produce the other one, so the flux ratio was set as 1:1.

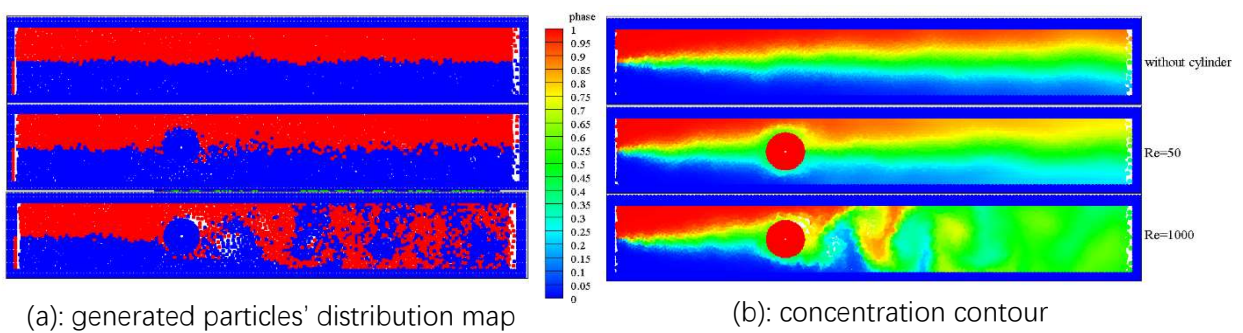

(a): generated particles' distribution map    (b): concentration contour

Fig. 14 Illustration of result without and with diffusion model

Fig.14 gives the particles and the concentration distribution of situation of non-cylinder, with cylinder and with cylinder and unsteady flow. Fig.14(a) could be considered as a passive mixer where non-diffusion happens. undoubtedly unsteady flow is very helpful for mixing and that's why the diffusion process can get accelerated when diffusion model is added to calculate the mass transfer phenomenon. According to that, the key to obtain well diffused fluid is to increase complicacy on flow. Simply a cylinder obstacle would dramatically result in better performance. The reasoning could go further by analysis as follows. The flow around a circular cylinder at Reynold number= 1000 periodically produce shedding vortex which strengthens the mass transfer. The periodical illustration is as fig.15 shows.

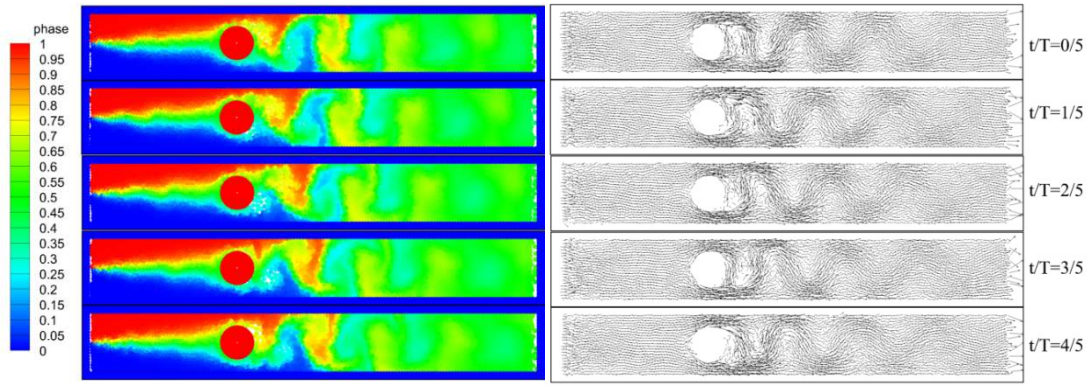

Fig. 15 Periodical illustration of concentration contour and vector

Due to the periodical effect, the outlet concentration distribution scatter map was draw in a 3D space with term of time added as fig. 16. The inhomogeneity reduced with time advancing comparing the initial moment. Similar fluctuation also occurred while the distribution finally centralized at an ideal level.

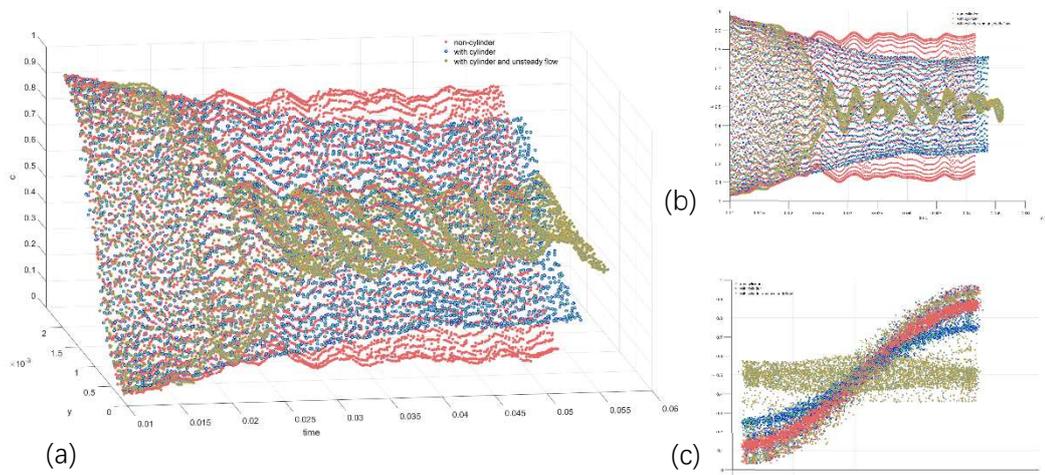

(a)    (b)    (c)

Fig. 16 3D scattering of outlet concentration. (a): the default view; (b): the front view; (c): the back view)

The periodical outlet average concentration data was analyzed by FFT (Fast Fourier Transform) to find the main frequency.

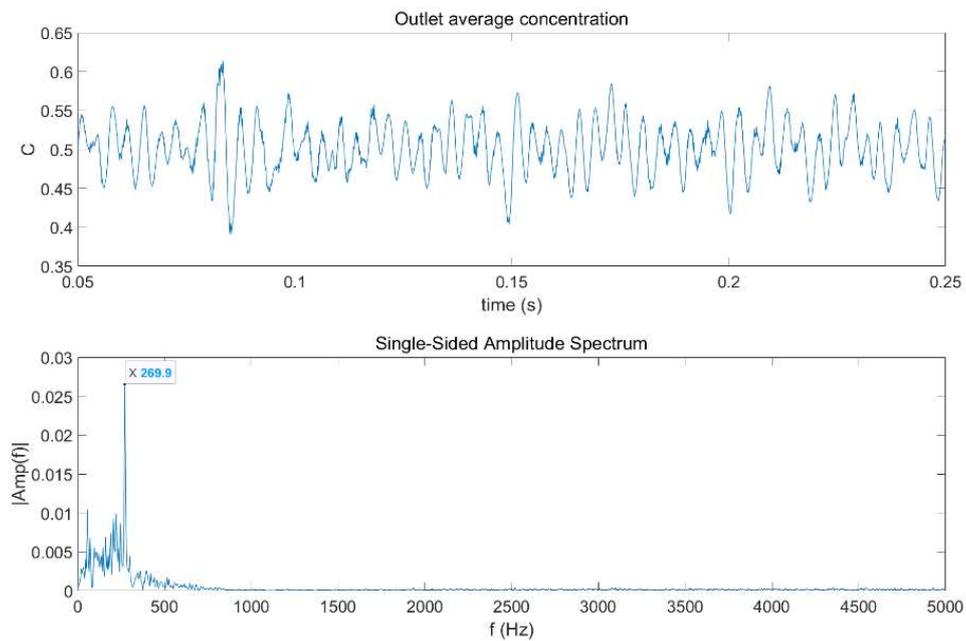

Fig. 17 FFT analysis of outlet average concentration

As fig.17 indicates, the main frequency of outlet average concentration is Fc=269.9 Hz, which is almost the same with the frequency of shedding vortex Fs=271 Hz (obtained by the flow period is 0.0037s). Therefore, the fluctuations of concentration could be considered as the result of shedding vortex effect.

It is known that shear flow will accelerate the progress of diffusion. In our method, particles were regarded as fluid assembles in Lagrangian description. Due to the lagrangian characters, the pathlines can be directly, correctly and conveniently obtained. In that case, the analysis of any part of fluid is able to be taken. Marked pathline a, b, c and d were extracted to analysis for representatives. Fig.18 gives pathlines of particles produced by inlet boundary condition at time= 0.05s. From the pathlines, the convection movement was strengthened after passing the cylinder where particle or part of fluid was taking part in furious mass transporting.

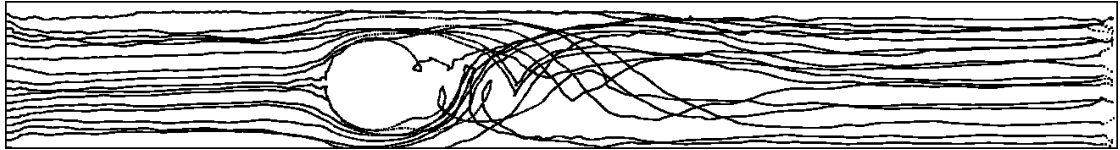

Fig. 18 Pathline of particles produced at time=0.05s and 4 extracted lines

The particle produced at 7/20 place of inlet line is showed as fig 18 a, the lifetime is 0.02561 second. The pathline length is 0.021325 m. Compared with the case without cylinder and unsteady flow, the lifetime is 70.7% longer and pathline length is 42.2% longer. While the particle produced at 13/20 place (fig 19 b) of the inlet line has a lifetime of 0.01971 second which is 31.4% longer than the without and a length of 0.021027 m which is 40.5% longer.

Table 1 Data of pathlines

| Pathline | a | b | c | d |
| --- | --- | --- | --- | --- |
| Lifetime(s) | 0.02561 | 0.01971 | 0.02071 | 0.02229 |
| Pathline length(m) | 0.021325 | 0.021069 | 0.021018 | 0.021096 |
| Prolonging ratio(%) lifetime/pathline | 70.7 / 42.2 | 31.4 / 40.5 | 38.1 / 40.1 | 48.6 / 40.7 |

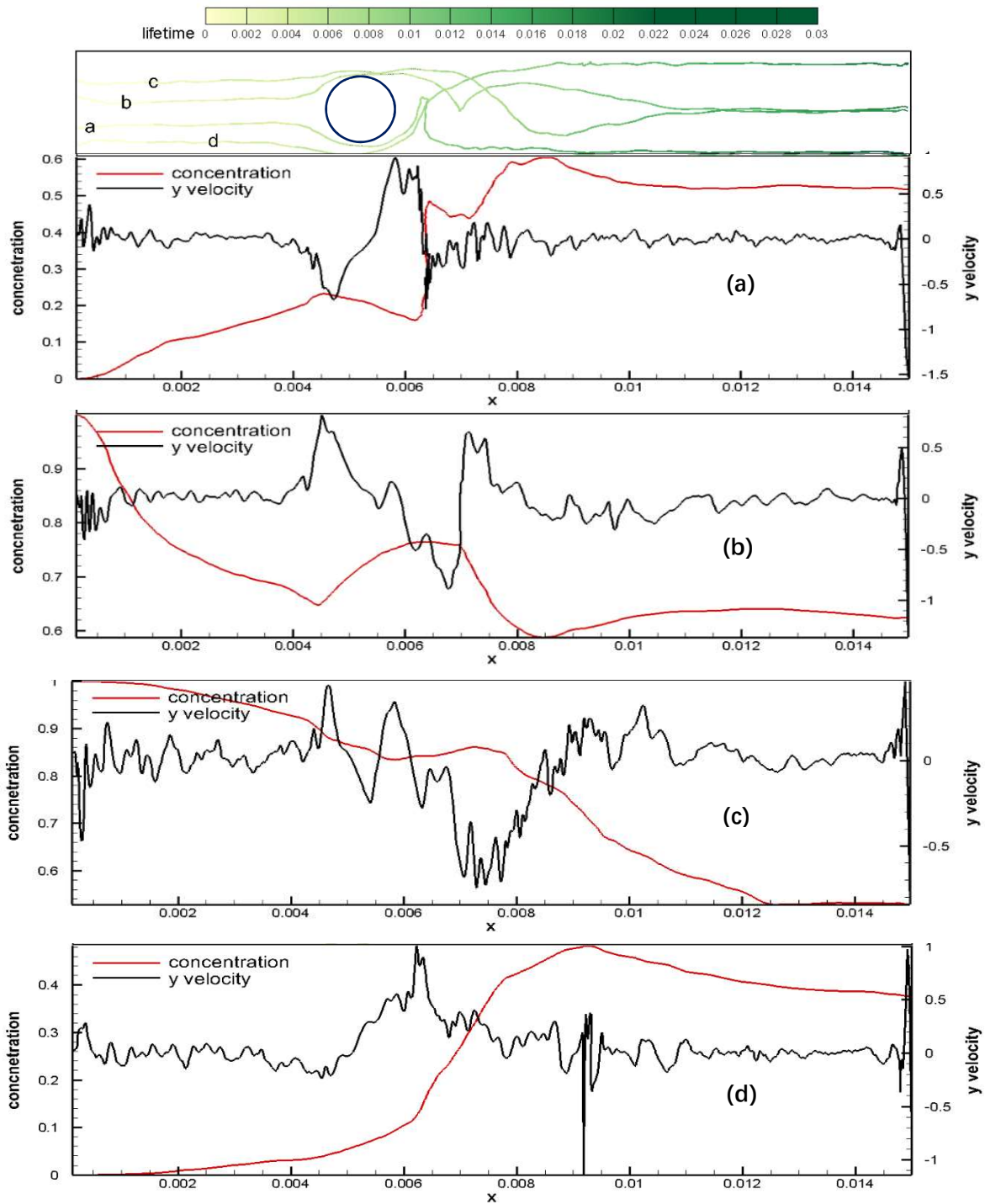

Fig. 19 Extracted pathlines with concentration, y velocity and lifetime variation (a, b, c and d are pathlines produced from different inlet line position)

It can not be simpily reasoned that the result is due to the prolonged lifetime and pathline, but it is certain that prolonging the lifetime and pathline do benefit the diffusion process. In other way the prolonging effect can be considered as a economic way to prolong the flow path without extending the length of tunnel. The concentration varies rapidly after the shear flow occur in fig.18. Due to the transportting movement, strongly mixing between the high-concentration and the low-concnetration parts occurs after passing the cylinder. The mass transfer variation can be pressented clearly by recording the concentration along the pathline. As illustrated in fig.19(a), concentration incured a violent climbing after the velocity had a climax at y direction, while in fig.18b, concnetration droped quickly due to passing a lowest point of velocity at y direction. Variation of concentration goes greater as the velocity varies. Before passing around the cylinder, fluctuation of velocity at y direaction happen around zero. Affected by passing around obstacle and shedding vortex, dramatical variation would happen and incurring rapid mixing or diffusion.

# Conclusion

Moving particle semi-implicit code was coupled with diffusion model based on Fick's $2^{nd}$ law. By defining particles as assembles of several component, it is convenient to calculate multi-component mass diffusion problem. The Laplacian operator shows most accuracy when set the radius of neighborhood particle searching as $2.5l0$. The validity of diffusion model of solving linear partial differential equation can be obtained in the comparison at two component diffusion problem. The flow involved case's validation was also performed by comparison with Fluent. The proper applications of boundary condition can result in ideal stagnation zone. By coupling diffusion model, free surface and moving boundary problem were simulated. The effective work in common is enlarging the interface area and incurring flow. Difference is how it works: Droplets' interface got enlarged by collision while moving boundary accomplished by pushing and bringing fluid. Moreover, the reason of incurring flow of droplets is surface tension while it was the solid part's movement that causing fluid flow. The outstanding enhancing effect on the diffusion process by unsteady flow was analyzed. It showed that shedding vortex could be the reason of concentration fluctuation. The main factor of how unsteady flow affecting diffusion process could be considered as the shear movement and its side effect: the prolonging of pathline and lifetime.